\documentclass[manuscript]{aastex}

\slugcomment{Not to appear in Nonlearned J., 45.}

\shorttitle{Spectro-photometry for Transit Spectroscopy}
\shortauthors{Matsuo et al.}

\begin{document}


\title{A New Concept for Spectro-photometry of Exoplanets with Space-borne Telescopes}


\author{Taro Matsuo\altaffilmark{1}, Satoshi Itoh\altaffilmark{1}, Hiroshi Shibai\altaffilmark{1}, Takahiro Sumi\altaffilmark{1}, and}
\affil{Department of Earth and Space Science, Graduate School of Science, Osaka University, 1-1, Machikaneyamacho, Toyonaka, Osaka 560-0043, Japan}
\author{Tomoyasu Yamamuro\altaffilmark{2}}
\affil{Optocraft, 3-16-8-101, Higashi Hashimoto, Midori-ku, Sagamihara, Kanagawa 252-0144, Japan}



\begin{abstract}

We propose a new concept for spectral characterization of transiting exoplanets with future space-based telescopes. This concept, called as densified pupil spectroscopy, allows us to perform high, stable spectrophotometry against telescope pointing jitter and deformation of the primary mirror. This densified pupil spectrometer comprises the following three roles: division of a pupil into a number of sub-pupils, densification of each sub-pupil, and acquisition of the spectrum of each sub-pupil with a conventional spectrometer. Focusing on the fact that the divided and densified sub-pupil can be treated as a point source, we discovered that a simplified spectrometer allows us to acquire the spectra of the densified sub-pupils on the detector plane-an optical conjugate with the primary mirror-by putting the divided and densified sub-pupils on the entrance slit of the spectrometer. The acquired multiple spectra are not principally moved on the detector against low-order aberrations such as the telescope pointing jitter and any deformation of the primary mirror. The reliability of the observation result is also increased by statistically treating them. Our numerical calculations show that, because this method suppresses the instrumental systematic errors down to 10 ppm under telescopes with modest pointing accuracy, next-generation space telescopes with more than 2.5$m$ diameter potentially provide opportunities to characterize temperate super-Earths around nearby late-type stars through the transmission spectroscopy and secondary eclipse.

\end{abstract}


\keywords{techniques: photometric --- techniques: spectroscopy --- planets and satellites: composition}



\section{INTRODUCTION}

Detection and spectral characterization of reflected light and thermal emission from an exoplanet is essential for understanding its atmospheric properties and compositions. The first successful spectral characterization of an exoplanet was reported by Charbonneau et al. (2002) through the primary transit spectroscopy of the exoplanet around HD 209458 with the STIS spectrograph of the Hubble Space Telescope. Following their success, Grillmair et al. (2007) and Richardson et al. (2007) respectively measured the emission spectra from HD 189733b and HD 209458b obtained with the Spitzer Infrared Spectrograph (IRS), thanks to the absence of an atmosphere and its stable environment compared to ground-based observatories. However, even observational data from space are still dominated by large instrumental systematic noise caused by image movement on the intra- and inter-pixel sensitivity variability due to telescope pointing jitter. For example, although the 3.6$\micron$ transit depth of the exoplanet HD 189733b has been measured by various groups (Tinetti et al. 2007; Ehrenreich et al. 2008; Beaulieu et al. 2008; D\`esert et al. 2009) with the same dataset acquired with Spitzer/IRAC, in the end no conclusion on the photometry of HD 189733b has been arrived at because of the inconsistency of their results. The discrepancy arises from fitting the instrumental systematic effects with a (non-) linear function assumed by each observer. Recently, Morello et al. (2014) applied a non-parametric statistical technique, called Independent Component Analysis (Waldmann 2012), to these datasets and then acquired a more reliable scientific output. 

Direct detection of exoplanets is more effective for characterization because the observational data are less affected by instrumental systematic noise than the transit spectroscopy. Marois et al. (2008) reported that an exoplanet orbiting the A-type star of HR8799 was directly imaged with Keck/Gemini 12 years after the first detection of an exoplanet. Snapshots of exoplanets were successfully obtained with other ground-based telescopes (e.g., Lagrange et al. 2010; Kuzuhara et al. 2013). The number of directly imaged exoplanets will rapidly increase thanks to the emergence of extreme adaptive optics with mature high-precision wavefront measurement and control techniques such as the Gemini Planet Imager (Macintosh et al. 2014) and the Subaru Coronagraphic Extreme AO (Jovanovic et al. 2015). However, detections and spectrum measurements of the terrestrial planets around G-type stars involve a more significant technical challenge because much higher contrast ratio values of $10^{-10}$ and $10^{-7}$ are required at visible and mid-infrared wavelengths. Even in those around M-type stars, the required contrast ratio values are still $10^{-8}$ and $10^{-5}$ respectively. Various approaches on direct imaging of the terrestrial planets have been proposed thus far: space-based coronagraph and occulter operating at visible wavelengths and space-based nulling interferometer operating at mid-infrared. Recently, a visible coronagraph instrument with a high-precision wavefront control mounted on the Wide-Field Infrared Survey Telescope-Astrophysics Focused Telescope Assets (WFIRST-AFTA) has been proposed as a technical demonstration of future dedicated visible coronagraph space missions for detection and characterization of the reflected light from Earth-like planets (Spergel et al. 2013). On the other hand, there is no way to detect and characterize the emission and the transmission spectra from the terrestrial planets and super-Earths because space-based nulling interferometers such as the Terrestrial Planet Finder Interferometer and Darwin have been indefinitely put on hold because of the significant technical challenges. However, detection and characterization of the emission spectrum are complementary to that of the reflected light and are essential for measurement not only of atmospheric composition but also of effective temperature and temperature-pressure profile. The remaining possibility for detection and characterization of the thermal emission is spectroscopy of the secondary eclipse with a more highly stabilized spectrophotometer mounted on a cryogenic telescope, compared to the IRS and IRAC of the Spitzer space telescope. In addition, the Transiting Exoplanet Survey Satellite (TESS) and the Planetary Transits and Oscillations of stars (PLATO) will be respectively launched in 2017 and 2022 and will then provide us with a number of smaller and cooler transiting exoplanets around nearby stars (Ricker et al. 2015; Rauer et al. 2014).

A general solution for high-precision photometry is to put a detector on the pupil plane, where it is an optical conjugate to the primary mirror, because the pupil photometry is immune to low-order wavefront errors such as tilt and defocus. Recently, Southworth et al. (2009) proposed high-precision photometry for measurement of planetary transits by telescope defocusing to minimize the telescope tracking error as well as saturation and flat-fielding of the charge-coupled device (CCD) detector and then achieved the level of the photometry down to 1 mmag even from ground. Although defocused spectroscopy by telescope defocusing is similar to the defocus photometry in that the above various systematic errors are reduced, the impact of the telescope pointing error on photometry still remains (Burton et al. 2015). One potential way to realize pupil spectrophotometry is to employ a superconducting detector such as Microwave Kinetic Inductance Detector (e.g., Mazin et al. 2012), which enables to measure the energy of an individual photon arriving at the detector, on the pupil plane. Note that, because such a superconducting detector should acquire only a single photon per the detector readout to accurately measure the energy of the photon, the resolving power of the detector decreases when observing bright stars. Another technique for high-precision photometry is to employ an orthogonal transfer CCD imager (OTCCD). The OTCCD measures the centroid of bright stars acquired with fast readout and then corrects low order tip-tilt adaptive optics (AO) by shifting the collected charge on the array during the science integration. As a result, the image quality corrected by the OTCCD is 0.1-0.15 arc-second in full-width half maximum (FWHM), corresponding to that acquired with a conventional AO system. Johnson et al. (2008) obtained photometry with 0.5 mmag precision per one data point, applying the OTCCD to measure the transit light curve. 

On the basis of this background, we propose a new method for transit spectroscopy with future space-borne telescopes. This new method, called densified pupil spectroscopy, realizes pupil spectrophotometry with a number of small telescopes aligned on an aperture plane such as the PLATO mission using a single large telescope. This method is similar to the lenslet-based integral field spectrograph such as TIGER (Bacon et al. 1995), which allows us to acquire the three-dimensional (two dimensional spatial and one dimensional spectral) information on astronomical objects, in that spectroscopy of the pupil light is performed. In other words, this concept is similar to a lenslet-based spectrograph that images the telescope pupil. This densified pupil spectroscopy has the several following advantages over a conventional spectrometer. First, because the detector plane, on which the spectra are observed, is an optical conjugate with the telescope primary mirror, the position of the acquired spectra are not principally changed against the telescope pointing jitter and any deformation of the primary mirror. Second, because the spectra are obtained over the entire detector, the scientific data can be reconstructed without defective pixels and pixels hit by cosmic rays. Third, many equivalent datasets can be acquired simultaneously. As a result, this concept can be easily applied to non-parametric data analysis such as coherent analysis (Swain et al. 2010; Waldmann et al. 2012) and independent component analysis (Waldmann et al. 2014) in order to reduce the remaining systematic noise. Thus, this method enhances optical and near-infrared transmission spectroscopy and mid-infrared secondary eclipse on space telescopes and even high altitude balloons with modest pointing accuracy. The proposed next-generation stable large space telescopes such as a Large UV/Optical/Near-Infrared Telescope (LUVOIR) and a Habitable Planet Imager (HabEx) potentially achieve photon noise below 10ppm, which enables to perform transmission spectroscopy of temperate terrestrial planets and super-Earths around late-type stars, thanks to the large collecting area. In addition, because planned or proposed cryogenic telescopes such as SPICA and CALISTO will provide us with a low background in mid-infrared wavelengths, application of this method to future cryogenic telescopes is a potential pathway to detection and characterization of the thermal emission from cooler super-Earths as shown in Section 3.

In this paper, we propose densified pupil spectroscopy for future space-borne telescopes. In Section 2, we give an overview of this concept and then describe its mathematical description based on Fresnel propagation. In Section 3, we evaluate the photometric stability of an optimized optical system for this concept and then discuss the capability for detection and characterization of thermal emissions from super-Earths with this system on a cryogenic space-borne telescope. 
 
\section{THEORY}
In this section, we give an overview of the proposed method and then present a mathematical description of wavefront propagation using the Fresnel diffraction equation to understand the nature and property of the method.  

\subsection{Overview}
This new concept for space transit spectroscopy, called densified pupil spectroscopy, achieves pupil spectrophotometry with a number of small telescopes aligned on an aperture plane such as the PLATO mission. This densified pupil spectroscopy allows us to perform highly stable spectrophotometry against telescope pointing jitter and deformation of the primary mirror instead of not having imaging capability. Figure 1 gives an overview of this concept. The densified pupil spectroscopy first divides the telescope aperture into a number of sub-pupils. Each sub-pupil is densified with two lens arrays. Here, focusing on the fact that the divided and densified sub-pupil can be treated as a point source, we discovered that a simplified spectrometer allows us to acquire the spectra of the densified sub-pupils on the detector plane, which is an optical conjugate with the primary mirror, by putting the divided and densified sub-pupils on the entrance slit of the spectrometer. The spectral resolution of this method is characterized by a diameter of the beam formed on the dispersive element, as discussed in Section 2.3. The beam diameter requires enlargement for higher spectral resolution. Thus, this concept realizes spectroscopy of the sub-pupils on the telescope aperture with simplified optics. This densified pupil spectroscopy has the following advantages compared to a conventional spectrometer in terms of the spectrophotometry. First, by producing an optical conjugation between a telescope aperture and a detector plane, this concept allows us to significantly reduce systematic noises due to telescope pointing jitter and distortion of the primary mirror, which prevent precise characterization of transiting exoplanets via the current space observatories such as the Spitzer space telescope and the Hubble Space Telescope. This concept also provides us with highly reliable scientific data by optically dividing the telescope aperture into a number of sub-apertures and simultaneously producing multiple equivalent spectra because the scientific data can be constructed without defective pixels and pixels hit by cosmic rays. In addition, the systematic noise can be further reduced through averaging of the multiple equivalent data in cases where the systematic noises are randomly added to the spectra. In stead of the averaging operation, various non-parametric data analysis techniques such as coherent analysis (Swain et al. 2010; Waldmann et al. 2012) and independent component analysis (Waldmann et al. 2014) are also applied to this method. Finally, thanks to division the telescope aperture into multiple small sub-apertures, any optical requirement on the primary mirror is mitigated. Thus, this method reduces the systematic noise, which is a major issue in current space transit spectroscopy, and potentially achieves observation performance limited not by the instrumental systematic noise but by shot noise. 

\subsection{Wavefront propagation}
In this section, we mathematically describe the wavefront propagation in the spectrometer of this densified pupil spectroscopy to understand its nature. From the standpoint of physical optics, this concept is mainly divided into two units: a pupil division/densifier unit and a conventional spectrometer. When the size of the densified sub-pupil formed by the former unit is close to the wavelength, the wavefront propagation in the spectrometer should not be written by geometric optics but by physical optics. In accordance with this consideration, the wavefront propagation in the spectrometer is mathematically described based on Fresnel propagation. The coordinate systems and parameters are set for this mathematical description as follows. The size of the entrance pupil is set to be $D$. The number of divisions along one dimension and the pupil densification are respectively defined as $n$ and $\eta$. The $n^{2}$ densified sub-pupils with size $\eta D/n$ are aligned in the re-imaged pupil plane, P2 in Figure 1. The P2 plane corresponds to the entrance slit of the spectrometer. The optics, composed of a collimator and camera optics in the spectrometer, produce an optical conjugate between the entrance slit and the detector plane. The coordinate systems in the entrance slit (P2), the plane just after the collimator optics (P3), the dispersive element plane (P4), the plane just before the camera optics (P5), and the detector plane (P6) are defined as $(x_{2}, y_{2})$, $(x_{3}, y_{3})$, $(x_{4}, y_{4})$, $(x_{5}, y_{5})$, and $(x_{6}, y_{6})$, respectively. The electric field in the coordinate system $(x_{i}, y_{i})$ is defined as $E_{i}(x_{i},y_{i})$. The collimator and camera elements, whose focal lengths are $f_{1}$ and $f_{2}$ and aperture functions are $A_{1}(x_{3},y_{3})$ and $A_{2}(x_{5},y_{5})$, are respectively placed at distances of $f_{1}$ and $2f_{1}+f_{2}$ from the P2 plane along the vertical direction of the entrance plane, $z$. The dispersive element, which is mathematically described by $M(x_{4},y_{4})$, is placed at a distance of $f_{1}$ from the collimator element along $z$. The detector is placed at a distance of $f_{2}$ from the camera optics.     

We first derive the electric field formed on the dispersive element for each densified sub-pupil of size $\eta D/n$ on the entrance slit of the spectrometer. As explained in Section 2.1, $n$ and $\eta$ should be designed such that the following approximation formula is satisfied:
\begin{equation}
f_{1} \gg \frac{(\eta D/n)^{2}}{\lambda}.
\end{equation}
According to Equation (1), the Fraunhofer diffraction pattern of the densified sub-pupil is formed on the plane (P3) through the collimator element: 
\begin{equation}
E_{3}(x_{3},y_{3}) = \frac{\exp(\frac{2 \pi i f_{1}}{\lambda})}{i \lambda f_{1}}A_{1}(x_{3},y_{3})\int dx_{2} \int dy_{2} E_{2}(\frac{x_{2}}{f_{1}}, \frac{y_{2}}{f_{1}})\exp \biggl(\frac{-2 \pi i (x_{2}x_{3}+y_{2}y_{3})}{\lambda f_{1}}\biggr).
\end{equation}
By using Fourier transform, the above equation can be written as
\begin{equation}
E_{3}(x_{3},y_{3}) = \frac{f_{1}\exp(\frac{2 \pi i f_{1}}{\lambda})}{i \lambda}A_{1}(x_{3},y_{3})FT\{E_{2}(\frac{x_{2}}{f_{1}},\frac{y_{2}}{f_{1}})\},
\end{equation}
where $FT\{h\}$ is the Fourier transform of a function $h$. The wavefront propagation from the plane P3 to the dispersive element can be described by the Fresnel diffraction equation. The electric field on the dispersive element is 
\begin{equation}
E_{4}(x_{4},y_{4}) = - \frac{\exp(\frac{4 \pi i f_{1}}{\lambda})}{\lambda^{2}}\int dx_{3} \int dy_{3} A_{1}(x_{3},y_{3}) \exp\biggl(\frac{\pi i \bigl((x_{3}-x_{4})^{2}+(y_{3}-y_{4})^{2}\bigr)}{\lambda f_{1}}\biggr)FT\{E_{2}(\frac{x_{2}}{f_{1}},\frac{y_{2}}{f_{1}})\}.
\end{equation}
The above equation can be written by a convolution as follows: 
\begin{eqnarray}
E_{4}(x_{4},y_{4}) = - \frac{\exp(\frac{4 \pi i f_{1}}{\lambda})}{\lambda^{2} f_{1}^{2}}\biggl[\exp\biggl(\frac{\pi i(x_{3}^{2}+y_{3}^{2})}{f_{1} \lambda}\biggr)*\biggl(A_{1}(x_{3},y_{3})FT\{E_{2}(x_{2},y_{2})\}\biggr)\biggr](x_{4},y_{4}),
\end{eqnarray}
where $*$ is the convolution operator. By using the convolution theorem and the following equation,
\begin{equation}
FT^{-1} \biggl\{ \exp(\frac{\pi i (x_{3}^{2}+y_{3}^{2})}{\lambda f_{1}}) \biggr\} = \frac{\lambda f_{1}}{2} \exp \biggl(-\frac{\pi i (x_{2}^{2}+y_{2}^{2})}{\lambda f_{1}}+\frac{\pi i}{2} \biggr),
\end{equation}
Equation (5) is rewritten as
\begin{equation}
E_{4}(x_{4},y_{4}) = - \lambda f_{1} \exp\biggl(\frac{4 \pi i f_{1}}{\lambda}+\frac{\pi i}{2}\biggr) FT\biggl\{ \exp\biggl(-\frac{\pi i(x_{2}^{2}+y_{2}^{2})}{\lambda f_{1}}\biggr) \biggl(FT^{-1}\{A_{1}(x_{3},y_{3})\}*E_{2}(\frac{x_{2}}{f_{1}},\frac{y_{2}}{f_{1}})\biggr) \biggr\}.
\end{equation}
By applying a large size collimator element to this system, $FT^{-1}\{A_{1}(x_{3},y_{3})\}\simeq \delta(\frac{x_{2}}{f_{1}})\delta(\frac{y_{2}}{f_{1}})$. Since the diameter of each densified sub-pupil on the entrance slit, $\eta D/n$, is fully smaller than the length of $f_{1}$, the electric field on the dispersive element is rewritten as 
\begin{equation}
E_{4}(x_{4},y_{4}) = - \lambda f_{1}\exp\biggl(\frac{4 \pi i f_{1}}{\lambda}+\frac{\pi i}{2}\biggr) FT\{E_{2}(\frac{x_{2}}{f_{1}},\frac{y_{2}}{f_{1}})\}.
\end{equation}
When Equation (1) is satisfied, the Fraunhofer diffraction pattern of the densified sub-pupil on the entrance slit is formed on the dispersive element. Comparing Equations (3) with (8), the Fraunhofer diffraction pattern is kept from the plane P3 just after the collimator optics to the dispersive element. As a result, a tilt of the phase in the densified sub-pupil, corresponding to the telescope pointing error, does not change the incident angle to the dispersive element but its irradiated area by the diffraction image. Thus, the spectrum formed on the detector plane is very stable against the telescope pointing jitter.

We next derive the electric field formed on the detector plane. The electric field just after the dispersive element is 
\begin{equation}
E_{4}^{'}(x_{4},y_{4}) = E_{4}(x_{4},y_{4})M(x_{4},y_{4}).
\end{equation}
The electric field formed just before the camera optics is described through the Fresnel diffraction equation as follows:
\begin{equation}
E_{5}(x_{5},y_{5}) = \frac{\exp(\frac{2 \pi i f_{2}}{\lambda})}{i \lambda f_{2}} \int dx_{4} \int dy_{4} E_{4}^{'}(x_{4},y_{4}) \exp \biggl(\frac{\pi i (x_{4}-x_{5})^{2} + (y_{4} - y_{5})^{2}}{\lambda f_{2}}\biggr)
\end{equation}
Since the electric field formed on the detector plane becomes the Fraunhofer diffraction pattern of the electric field just before the camera optics,
\begin{eqnarray}
E_{6}(\frac{x_{6}}{f_{2}},\frac{y_{6}}{f_{2}}) = - \frac{\exp(\frac{4 \pi i f_{2}}{\lambda})}{\lambda^{2} f_{2}^{2}} \int dx_{5} \int dy_{5} A_{2}(x_{5},y_{5}) \exp \biggl(\frac{2 \pi i (x_{5}x_{6}+y_{5}y_{6})}{\lambda f_{2}} \biggr) \nonumber \\
\times \int dx_{4} \int dy_{4} E_{4}^{'}(x_{4},y_{4}) \exp \biggl(\frac{\pi i \bigl((x_{4}-x_{5})^{2}+(y_{4}-y_{5})^{2}\bigr)}{\lambda f_{2}}\biggr)  
\end{eqnarray}
By describing it with a convolution, the above equation can be written as follows:
\begin{eqnarray}
E_{6}(\frac{x_{6}}{f_{2}},\frac{y_{6}}{f_{2}}) = - \frac{\exp(\frac{4 \pi i f_{2}}{\lambda}+\frac{\pi i}{2})}{\lambda^{2} f_{2}^{2}} \biggl[ FT^{-1}\biggl\{ \exp(-\frac{\pi i(x_{4}^{2}+y_{4}^{2})}{\lambda f_{2}}) \biggr\} \nonumber \\ 
\times FT^{-1}\{E_{4}^{'}(x_{4},y_{4})\} \biggr] * FT^{-1}\{A(x_{5},y_{5})\}.
\end{eqnarray}
Given that the size of the camera element is fully large as well as the collimator, $FT^{-1}\{A(x_{5},y_{5})\}\simeq \delta(\frac{x_{6}}{f_{2}}) \delta(\frac{y_{6}}{f_{2}})$. By inserting Equations (8) and (9) into the above equation, the electric field on the detector plane becomes
\begin{equation}
E_{6}(\frac{x_{6}}{f_{2}},\frac{y_{6}}{f_{2}}) = - \frac{f_{1}}{f_{2}}\exp\biggl(\frac{4 \pi i (f_{1}+f_{2})}{\lambda}\biggr)\biggl[FT^{-1}\{M(x_{4},y_{4})\} * E_{2}(\frac{x_{2}}{f_{1}},\frac{y_{2}}{f_{1}})\biggr].
\end{equation}
The intensity distribution on the detector plane is simply expressed as a convolution of the densified sub-pupil on the entrance slit, $E_{2}(\frac{x_{2}}{f_{1}},\frac{y_{2}}{f_{1}})$, with the Fourier transform of the dispersive element function, $FT\bigl\{M(x_{4},y_{4})\bigr\}$. Consequently, the position of the dispersive element between the collimator and the camera elements does not affect the dispersion property.

Finally, the above mathematical descriptions are extended to the case of $n^{2}$ densified sub-pupils, instead of each one individually. The electric field on the entrance slit, $E_{2}(\frac{x_{2}}{f_{1}},\frac{y_{2}}{f_{1}})$, can be expressed as a summation of each densified sub-pupil:
\begin{equation}
E_{2}(x_{2},y_{2}) -> \sum_{i}^{n^{2}} E_{2,i}(\frac{x_{2}}{f_{1}},\frac{y_{2}}{f_{1}}).
\end{equation} 
As a result, an interference pattern is formed on the dispersion element:
\begin{equation}
E_{4}(x_{4},y_{4}) = - \lambda f_{1}\exp\biggl(\frac{4 \pi i f_{1}}{\lambda}+\frac{\pi i}{2}\biggr) FT\biggl\{\sum_{i}^{n^{2}}E_{2,i}(\frac{x_{2}}{f_{1}},\frac{y_{2}}{f_{1}})\biggr\}.
\end{equation}  
On the other hand, because each spectrum is separated on the detector according to Equation (13), the dispersive property is the same as for the case of a single sub-pupil. We continue to consider the case of a densified sub-pupil to understand the nature of this concept in the ensuing discussion.

\subsection{Spectral resolution}
On the basis of the analytical description of the wavefront propagation in Section 2.2, we derive the spectral resolution of this concept for the case where a grating is adopted as the dispersive element. When the pitch of the grating and the number of the cycles are respectively $g$ and $N$, the grating function, $M(x_{4},y_{4})$, can be represented as  
\begin{equation}
M(x_{4},y_{4})=G(x_{4},y_{4})*\biggl\{ \sum_{k=0}^{N-1}\delta(x_{4}-gk)\exp(ik\Phi) \biggr\},
\end{equation}
where $G(x_{4},y_{4})$ is the sub-pupil function in each cycle and $\Phi$ is the phase difference between two adjacent cycles. Based on Equation (13), the dispersion property of this concept is characterized by the Fourier transform of the grating function, $FT\{M(x_{4},y_{4})\}$:
\begin{equation}
FT\{M(x_{4},y_{4})\}=\lambda^{2}\delta \biggl(\frac{y_{6}}{f_{2}}\biggr)\frac{sin\biggl(\frac{N(\frac{2\pi g x_{6}}{\lambda f_{2}}+\Phi)}{2}\biggr)}{sin\biggl(\frac{(\frac{2\pi g x_{6}}{\lambda f_{2}}+\Phi)}{2}\biggr)}FT^{-1}\{G(x_{4},y_{4})\},
\end{equation}
where $FT^{-1}\{G(x_{4},y_{4})\}$ represents the efficiency of the grating. From the above equation, the maximum displacement of the spectrum element on the detector is determined by
\begin{equation}
\frac{N(\frac{2\pi g x_{6}}{\lambda f_{2}}+\Phi)}{2}=m\pi,
\end{equation}
where $m$ represents the order of the the dispersive element. As a result, the relation between the image displacement on the detector plane and the wavelength of the spectrum element is derived as follows: 
\begin{equation}
\frac{g}{f_{2}}dx_{6}=md\lambda.
\end{equation}
Because the minimum sampling interval of each spectrally resolved component on the detector plane corresponds to the pupil size, $d_{p}$, as shown in Equation (13), the spectral resolution of this concept is
\begin{equation}
\frac{\lambda}{d\lambda}= \frac{m}{2} \frac{2 \lambda f_{2} / d_{p}}{g}.
\end{equation}
Because the number of the grating cycles included in the Fraunhofer pattern, $N_{PSF}$, is 
\begin{equation}
N_{PSF}=\frac{2 \lambda f_{2}}{gd_{p}},
\end{equation}
the spectral resolution is characterized by $m$ and $N_{PSF}$. Based on the above consideration, an increase is required in the number of divisions of the pupil along the pupil, $n$, and the densification of each pupil, $\eta$ for higher spectral resolution. Since the pupil size, $d_{p}$, and the pitch of the grating, $g$, cannot be reduced down to $\lambda$, the maximum spectral resolution, $R_{max}$, is
\begin{equation}
R_{max}\equiv\biggl\{\frac{\lambda}{d\lambda}\biggr\}_{max}<m\frac{f_{1}}{\lambda}.
\end{equation}

\subsection{Calculation of amplitude profile}
To investigate whether the above analytical descriptions on wavefront propagation are satisfied with actual parameters, we derive one-dimensional electric fields on the dispersive element and the detector plane based on the Fresnel diffraction equation. By assuming that aplanatic lenses are applied to the collimator and camera optics, the phase function given by passing through a lens of focal length $f$ is defined as
\begin{equation}
U(x,y)=\exp\biggl(-\frac{2 \pi i (r^{'}-f)}{\lambda}\biggr),
\end{equation}
where $(x,y)$ is the coordinate system on the lens and $r^{'}$ is the distance between the focal point of the lens and the position on the lens, $(x,y)$. The $r'$ is approximated by
\begin{equation}
r'\simeq f+\frac{x^{2}+y^{2}}{2f}.
\end{equation}
By using the Fresnel diffraction equation and Equations (22) and (23), the complex amplitudes on the various planes from P3 to P6 can be analytically calculated. Figure 2 shows the overall view of the optical system and the absolute values of the complex amplitudes on the entrance slit of the spectrometer, on the dispersive element, and on the detector plane. In this calculation, the parameters are set as follows. A single densified sub-pupil with diameter of 0.4$mm$ is put on the entrance slit for simplicity. The source is monochromatic light at a lambda of 0.01$mm$. The focal lengths of both the collimator and the camera optics have the same parameter of 100$mm$. The dispersive element is placed at distance of 100$mm$ from the collimator lens. The left side of Figure 2 shows the modulus of the complex amplitude on the entrance slit. Since the diameter of the densified sub-pupil is much smaller than the focal length of the collimator lens, the Fraunhofer diffraction pattern of the electric field on the entrance pupil is formed as shown in Equation (3). The center of Figure 2 shows the modulus of the complex amplitude on the dispersive element. The Fraunhofer diffraction pattern is kept in the propagation from P3 to P4 as a parallel light, as explained in Section 2.2. Consequently, as described in Equations (20) and (21), the spectral resolution of this method is determined by the grating cycles included in the Fraunhofer pattern, $N_{PSF}$. The right side of Figure 2 shows the absolute value of the complex amplitude on the detector plane. The densified sub-pupil on the entrance slit is re-imaged on the detector but the sub-pupil is slightly affected by spatial filtering due to the finite aperture of the optics. Thus, the mathematical description derived thus far is satisfied without any approximations. Note that the sizes of the collimator lens, the grating, and the camera lens should be ten times larger than Airy disk formed by each sub-pupil to avoid the spatial filtering of the optical elements.

\subsection{Partial occultation of target star with field stop}
This method principally liberates the photometry from the low-order aberrations such as the telescope pointing jitter and any deformation of the primary mirror as discussed in Section 2.1. However, the telescope pointing jitter affects the photometric stability in real systems for the following reason. Since the field of view should be limited by a field stop, which is a circular mask put on the image plane, to avoid contamination of background stars, the field stop occults the wings of the point spread function formed the target star. The partial occultation of the target star leads to degradation of the photometric stability under the pointing jitter of the telescope, $\theta_{jitter}$, during the transit observation. We estimated the degradation of the photometric stability due to the partial occultation of the point spread function formed on the image plane with the field stop through numerical simulations. Figure 3 shows the photometric stabilities as a function of the radius of the field stop for three different values of $\theta_{jitter}=(0.1\lambda/D, 0.5\lambda/D, 1\lambda/D)$. The photometric stabilities for all values of the pointing jitter are more improved as the mask radius increases. The upper limit of the photometric stability, $\Delta L_{mask}$, approximately obeys the following equation:
\begin{equation}
\Delta L_{mask} \leq \alpha (\theta_{jitter}) \theta_{mask}^{-2.3},
\end{equation}
where $\theta_{mask}$ is the mask radius in unit of $\lambda/D$ and $\alpha$ represents a constant as a function of the pointing jitter, $\theta_{jitter}$. $\alpha$ takes 0.0035 and 0.05 for the pointing jitters of $0.1\lambda/D$ and more than $0.5\lambda/D$, respectively. Based on this estimation, the radius of the field stop should be larger than $12.5\lambda/D$ for $\theta_{jitter}=0.1\lambda/D$ and 40 for $\theta_{jitter} > 0.5\lambda/D$ to achieve a photometric stability down to 10ppm. In other words, large pointing jitter easily leads to contamination of background stars because of requirement of a field stop with a large radius. Figure 4 shows the cumulative number density of galactic stars in the N band for various galactic coordinates, $(b,l)$, based on the calculation procedure derived by Konishi et al. (2015). The dashed black line in Figure 4 indicates the number density corresponding to that which a galactic star is at least contaminated within a field of view of 10 arc-second in radius. In other words, it is difficult to achieve a high photometric stability down to 10 ppm for the galactic plane, $b=0$, in case of the field of view of 10 arc-second in radius. Thus, as the pointing jitter increases, the observable region is more restricted.   
 
\section{PERFORMANCE}
In this section, we derive an analytical expression for the photometric stability of this pupil densified spectroscopy due to an image movement on the inter- and intra-pixel sensitivity variations and then evaluate its expected performance with an optimized optical system for this concept. Based on this performance, we discuss characterization of exoplanets through transmission spectroscopy and secondary eclipse with this system mounted on a cryogenic space telescope with 2.5$m$ diameter. 

\subsection{Analytical expression of photometric stability}
The photometric stability of this pupil densified spectroscopy is limited by image movement on the detector, which has inter- and intra-pixel sensitivity variations, given that the telescope defocus is not changed during the transit observation thanks to the thermally stable environment provided by the cryogenic telescope. We first consider the impact of the intra-pixel sensitivity variation on the photometric stability. When the position of the central gravity of a sub-pupil is $(x_{c},y_{c})$ on the detector plane, the observed luminosity of the sub-pupil, $L(x_{c},y_{c})$, is simply written as
\begin{equation}
L(x_{c},y_{c})=\int dx_{6} \int dy_{6} i(x_{6}+x_{c},y_{6}+y_{c}) \xi(x_{6},y_{6}),
\end{equation}
where $i(x_{6},y_{6})$ and $\xi(x_{6},y_{6})$ are the intensity profile of the sub-pupil on the detector and the pixel sensitivity with the intra-pixel sensitivity variation as a function of the position of the detector plane, respectively. In a case where the sub-pupil moves on the detector plane by $(\delta x_{c}, \delta y_{c})$, the luminosity can be rewritten as: 
\begin{equation}
L(x_{c}+\delta x_{c},y_{c}+\delta y_{c})=[i(x_{6},y_{6})*\xi(x_{6},y_{6})](x_{c}+\delta x_{c},y_{c}+\delta y_{c}).
\end{equation}
Here, on the basis of the sub-pixel sensitivity measurements of near-infrared detectors by Barron et al. (2006), the intra-pixel sensitivity is characterized by a convolution of a box function with a width of the pixel size, $p$, and the lateral charge diffusion factor. We approximate the lateral charge diffusion factor as a two-dimensional Gaussian function for simplicity. Assuming that all pixels have the same intra-pixel sensitivity, we can give the intra-pixel sensitivity variation as
\begin{equation}
\xi(x_{6},y_{6})=\sum_{k,l}\delta(x_{6}-kp, y_{6}-lp) * \biggl[\exp\biggl(-\frac{x^{2}+y^{2}}{(0.1p)^{2}}\biggr)*\biggl(rect(\frac{x}{p})rect(\frac{y}{p})\biggr)\biggr](x_{6},y_{6}),
\end{equation} 
where $p$ and $(k,l)$ represent the pixel pitch of the detector array and the two-dimensional pixel number, respectively, and the $rect$ function is defined as follows:
\begin{eqnarray}
\left\{
\begin{array}{l}
rect(x)=1 \ \ (|x|\leq1) \\
rect(x)=0 \ \ (otherwise)
\end{array}
\right.
\end{eqnarray}
By using Equations (28) and (29), the luminosity becomes
\begin{equation}
L(x_{c}+\delta x_{c},y_{c}+\delta y_{c})=\sum_{k,l}i^{'}(kp+\delta x_{c}, lp+\delta y_{c}),
\end{equation} 
where
\begin{equation}
i^{'}(x_{6},y_{6})=i(x_{6},y_{6})* \biggl[\exp\biggl(-\frac{x^{2}+y^{2}}{(0.1p)^{2}}\biggr)*\biggl(rect(\frac{x}{p})rect(\frac{y}{p})\biggr)\biggr](x_{6},y_{6}).
\end{equation} 
When the diameter of the sub-pupil on the detector, $\frac{\eta D}{n}$, is much larger than the pixel pitch, $p$, the intensity profile of $i^{'}(x_{6},y_{6})$ is constant except for both edges of the sub-pupil. In other words, the intra-pixel sensitivity variation has little influence on the luminosity. Otherwise, the intensity profile of $i^{'}(x_{6},y_{6})$ is affected by the intra-pixel sensitivity. On the basis of the above considerations, the photometric variation caused by an image movement on a detector with the intra-pixel sensitivity variation, $\Delta L_{intra}$, is finally described as
\begin{equation}
\Delta L_{intra} \equiv \ \frac{L(x_{6}+\delta x_{6},y_{6}+\delta y_{6})-L(x_{6},y_{6})}{L(x_{6},y_{6})}=\frac{\sum_{k,l}i^{'}(kp+\delta x, lp+\delta y)}{\sum_{k,l}i^{'}(kp,lp)}-1.
\end{equation}  
The above equation can be applied both to the photometric stability of this densified pupil spectroscopy and also to that of a conventional spectroscopy. 

Next, we analytically describe the impact of an image movement on a detector with an inter-pixel sensitivity variation. Assuming that the travel distance of the image is pixel-level, the luminosity of a sub-pupil can simply be written as a summation of the luminosity per pixel instead of an integration of pixels. When the sub-pupil extends across $N$ pixels along the two axes, the luminosity of the sub-pupil positioned at pixel number of $(l,m)$ is
\begin{equation}
L_{l,m}=i_{pixel}\sum_{l=-\frac{N}{2}}^{\frac{N}{2}}\sum_{m=-\frac{N}{2}}^{\frac{N}{2}}\xi_{l,m},
\end{equation}  
where $i_{pixel}$ and $\xi_{l,m}$ represent the luminosity per one pixel and the pixelized inter-pixel sensitivity variation described as a function of the pixel number, $(l,m)$, respectively. The photometric variation limited by the image motion on the inter-pixel sensitivity variation, $\Delta L_{inter}$, is
\begin{equation}
\Delta L_{inter} \equiv \ \frac{L_{l+\delta l,m+\delta m}-L_{l,m}}{L_{l,m}}\simeq \frac{\sum_{l=-\frac{N}{2}}^{\frac{N}{2}}\sum_{m=-\frac{N}{2}}^{\frac{N}{2}}(\xi_{l+\delta l,m+\delta m}-\xi_{l,m})}{N^{4}}.
\end{equation} 
Using the operating variance of Equation (34), $V(\Delta L_{inter})$, the standard deviation of the photometric stability is calculated as 
\begin{equation}
\sigma_{inter} \equiv \sqrt{V(\Delta L_{inter})} = \frac{\sqrt{2N_{motion}}\sigma_{\xi}}{N^{2}},
\end{equation} 
where $N_{motion}$, and $\sigma_{\xi}$ are respectively the number of pixels newly irradiated or not newly irradiated by the image motion and the standard deviation of the inter-pixel sensitivity variation. The above equation can be extended to the case of sub-pixel movement of the image and then $N_{motion}$ becomes not an integer but a real number. As $N$ increases, the impact of the inter-pixel sensitivity on the photometric stability decreases.   

In conclusion, the photometric stability limited by an image motion on the detector due to the telescope pointing jitter can be characterized by the sampling number of the sub-pupil, $N$, the number of the pixels newly irradiated or not newly irradiated by the image motion, $N_{motion}$, and the inter-pixel sensitivity variation. The size of the sub-pupil on the detector should be much higher than the pixel pitch, $p$, in order to suppress the instrumental systematic noises caused by the image motion on the intra- and the inter-pixel sensitivity variations. In contrast, the signal from the transiting planet is dominated by the detector noise such as dark current and readout noise. The sampling number of the sub-pupil on the detector should be optimized by a balance between the instrumental systematic noise and the detector noise.

\subsection{An optimized optical design for this concept}
In this section, we present an optimal design for the pupil densified spectroscopy concept. The diameter of the telescope primary is set to 2.5$m$. The observation wavelength range is set in the range 10 to 20$\micron$ for observation of thermal emissions from cooler planets less than 1000$K$. The spectral resolution at 10$\micron$ is 100 for measurement of the atmospheric compositions. Owing to the relation between the sampling number and the instrumental systematic errors, the optical system is designed such that the sampling number of the spectrally resolved sub-pupil is eight on the detector. The overall view of the optical system combined with the telescope is shown in Figure 5. The field of view is restricted with an aperture mask at the telescope's focal point, as described in Section 2.5. Given that the pointing jitter of the telescope is 0.1 arc-second rms in total, which approximately corresponds to requirement on the pointing stabilities of SPICA\footnote{ESA's CDF study report: Assessment of next generation cryogenic infrared telescope (reference number: CDF-152(A)}, the radius of the aperture mask is set to 10 arc-second for achieving the high photometric stability down to 10ppm based on Equation (25). Based on the number density of the galactic stars shown in Figure 4, the observable region for this system is limited to high galactic latitude of $|b|>30$. The beam diameter collimated by the first two lenses after the telescope focus, $D$, is 10$mm$. The collimated beam is divided into 20 sub-pupils by a mask positioned on the output pupil. The figure of the mask is same as that shown in Figure 1. After the two microlens arrays densify each sub-pupil with a densification factor of 10, densified sub-pupils with a diameter of 200$\micron$ are formed on the output pupil, corresponding to the entrance slit of the spectrometer. The light beam of one sub-pupil positioned at the center of the entrance slit is drawn in the spectrometer of Figure 4. After the beam collimated by the two lenses with an focal length of 75$mm$ is reflected by a reflection grating with 20 cycles per 1$mm$, the five lenses with an effective focal length of 75$mm$ form the spectrum of the sub-pupil on the detector with a pixel format of 1000 x 1000 and a pixel pitch of 25$\micron$. The detector applied to this system is a Si:As Impurity Band Conductor array for the mid-infrared wavelength range. The total of the optical throughput, including in the detector quantum efficiency and the grating efficiency, is set to 30$\%$ for the latter calculation. Figure 6 shows the footprint on the detector plane. Figure 7 shows the spot diagrams on the detector plane for various positions and wavelengths of the sub-pupils. All of the lenses in this optical system are designed such that the image enlargements due to the aberrations for all sub-pupils at 10 to 20$\micron$ are much smaller than the size of each sub-pupil formed on the detector plane. As a result, the spectra acquired on the detector are expected to be stabilized against pointing jitter.

\subsection{Expected performance on a cryogenic space-borne telescope}
On the basis on the derived analytical expression on photometric stability, we estimate the photometric stability of this optical system on a cryogenic telescope cooled to less than 10$K$, in which the thermal background from the telescope is negligible. Then, we compare the estimated performance with the depth of the secondary eclipse of a temperate super-Earth with a radius two times that of Earth around various spectral types of stars. We also consider transmission spectroscopy of an Earth-twin. Table 1 shows the average and worst values of the image motions on the detector in the case of a telescope pointing jitter of 0.1 arc-second rms in total. As predicted by the theory underlying this concept, the image motion on the detector due to the pointing jitter is extremely suppressed. Based on Equations (32) and (35), the photometric stability of this system can be calculated. However, the flat-fielding accuracy of the Si:As detector is unknown in the era of SPICA and CALISTO. Deming et al. (2009) applied to the flat-fielding error in Spitzer/IRAC of 0.4\% to calculate the photometric accuracy in the Mid-Infrared Instrument (MIRI) of JWST, assuming that the flat-fielding error in JWST/MIRI is not improved from the Spitzer space telescope. On the other hand, according to the Explanatory Supplement to the Wide-field Infrared Survey Exploler (WISE) all-sky data release products compiled by Cutri et al. (2013), the flat-fielding accuracy of the Si:As detector is 0.07\%. In this paper, we use 0.07\% as the flat-fielding error. Table 2 shows the mean and worst values of the instrumental systematic error caused by the pointing jitter in case of the flat-fielding accuracy of 0.07\%. Since the systematic noise error due to the image motion on the intra-pixel sensitivity variation is down to a level of $10^{-8}$ thanks to the large sampling number of the image, the systematic noise error is dominated by the inter-pixel sensitivity variation. Thus, the photometric stability is better than 10ppm. 

Next, we compare the photometric stability at 10$\micron$ with the photon noise, the detector noise, and exozodiacal light with the depths of the secondary eclipses of super-Earths around stars with various effective temperatures from 3200 to 5500$K$. The parameters used for this calculation are compiled in Tables 3 and 4. In order to estimate the depths of the secondary eclipses around various type of stars, we apply a relation between the star radius and the effective temperature, which was measured by Boyajian et al.(2012) with a stellar interferometer, as shown in Table 4. The effective temperature of the super-Earth is fixed to 300$K$. Given that the bond albedo of the planet is 0.3, the semi-major axis of the planet with an effective temperature of 300$K$ can be calculated. The transit durations of the planets around various type of stars are also estimated in the case of the impact parameter of $b=0.0$ based on the following relationship: 
\begin{equation}
T_{dur} \simeq \frac{P R_{*}}{\pi a},
\end{equation} 
where $P$ and $a$ are the period and the semi-major axis of the planet, respectively, and $R_{*}$ is the stellar radius. The semi-major axes and the transit durations of the planet with an effective temperature of 300$K$ around various types of stars are compiled in Table 4. The spectra of both the central star and the planet are taken to be blackbodies. The distance of this system is fixed at 10$pc$ because such a transiting super-Earth in the M dwarf habitable zone exists within about 10$pc$ at high confidence level, according to Dressing and Charbonneau (2015). In terms of the detector noise, we refer to the measurement values of a Si:As detector for the Mid-Infrared Instrument (MIRI) of the James Webb Space Telescope in Ressler et al. (2008). The dark current and the readout noise are set to 0.17$e^{-}/s/pixel$ and 14$e^{-}$, respectively. The exozodiacal lights around G- and A-type stars are recently modeled by Kennedy et al. (2015) for observations of the Large Binocular Telescope Interferometer (LBTI) based on the COBE/DIRBE observational results. According to this paper, the flux of the zodiacal light around a solar-type star is 100$\mu Jy$. The total integration time equals to multiplication of the transit duration and the number of the eclipses for three years, which is the mission lifetime of SPICA. Figure 8 compares this system performance with a requirement on 1-$\sigma$ detection of the secondary eclipse of a temperate super-Earth and its atmosphere of 20$km$ scale height through the transmission spectroscopy  around various types of stars at 10$\micron$. We also consider the performance of the conventional spectrometer, which has 8 pixel sampling per one spectral channel in the same manner as the new system. Thanks to the high stability against the pointing jitter in this system, the photometric stability can be much improved from that of the conventional one. The systematic noise is also comparable with the random noise in this observation. Comparing to the depths of the secondary eclipses of the temperate super-Earths around various types of the stars, this system potentially provides us with an opportunity of spectral characterization of the emissions from the temperate super-Earths around nearby late-type stars. In addition, this system can resolve the atmospheric thickness of 20$km$, which corresponds to the thicknesses of the ozone and the carbon dioxide layers in the Earth atmosphere (Kaltenegger \& Traub 2009), through the transmission spectroscopy with a spectral resolution of 20. According to Barstow et al. (2016), since the terrestrial planets with Venus-type and Earth-type around a M5 star may be distinguished through investigation the presence or absence of additional absorption of the ozone molecules around 10$\micron$. However, because the frequency spectrum of the pointing jitter is unknown at this time, only the upper limit on the partial absorption of the target star with the field stop is derived. The final performance on the systematic noise error is expected to settle into the gray shaded area shown in Figure 8. We also note that, because all systematic noises such as the detector gain variability and calibration error are not treated in this calculation, the above prediction on the science output is tentative. As the next step, we will investigate how to de-correlate the other systematic errors from the multiple spectra observed on the detector.

\section{CONCLUSION}
We proposed a new method, called densified pupil spectroscopy, for spectral characterization of transiting exoplanets. This densified pupil spectrometer consists of the following three roles: division of a pupil into a number of sub-pupils, densification of each sub-pupil, and acquisition of the spectrum of each sub-pupil with a conventional spectrometer. Focusing on a fact that the divided and densified sub-pupil can be treated as a point source, we found that a simplified spectrometer allows us to acquire the spectra of the densified sub-pupils on the detector plane, which is an optical conjugate with the primary mirror, by putting the divided and densified sub-pupils on the entrance slit of the spectrometer. The proposed methods results in several advantages compared to the conventional spectrometer. First, the detector plane, on which the spectra are observed, can be an optical conjugate with the telescope primary mirror. As a result, the position of the acquired spectra are not principally changed against the telescope pointing jitter and any deformation of the primary mirror. Second, the scientific data can be reconstructed without defective pixels and pixels hit by cosmic rays. Third, because many equivalent datasets can be acquired simultaneously, this concept can be easily applied to non-parametric data analysis such as coherent analysis and independent component analysis in order to reduce the remaining systematic noise. We analytically described the nature of this method and then evaluated the photometric performance of an optimal optical system designed in accordance with this concept. According to our numerical calculations, the instrumental systematic noise caused by the pointing jitter of 0.1 arc-second rms in total can be reduced down to 10 ppm. Compared to the requirement on 1-$\sigma$ detection of the secondary eclipse of a temperate super-Earth and its atmosphere of 20$km$ scale height, corresponding to the thicknesses of the ozone and the carbon dioxide layers in the Earth atmosphere, this system potentially characterizes the temperate super-Earths with thin atmosphere around nearby late-type stars through the transmission spectroscopy and secondary eclipse. As the next step, we plan to treat all systematic errors such as detector gain and calibration error and then investigate whether application of the non-parametric data analysis to this system can de-correlate them.  

\acknowledgments

We are sincerely grateful to Keigo Enya, Hiroaki Imada, Enzo Pascal, Peter Ade, and Matt Griffin for providing us possibilities of further development of this concept from both scientific and technical sides. We also thank Giovanna Tinetti and Ingo Waldmann for useful discussion towards establishment of an optimal calibration strategy for this concept. We acknowledge Hajime Kawahara and Takayuki Kotani for checking the photon-noise sensitivity and the depth of the secondary eclipse in our calculation. We also appreciate the calculation given by Konishi Mihoko for estimation of background stars. Finally, we would like to express our appreciation to an anonymous referee for valuable comments on this manuscript. 

\begin{figure}
\epsscale{.80}
\plotone{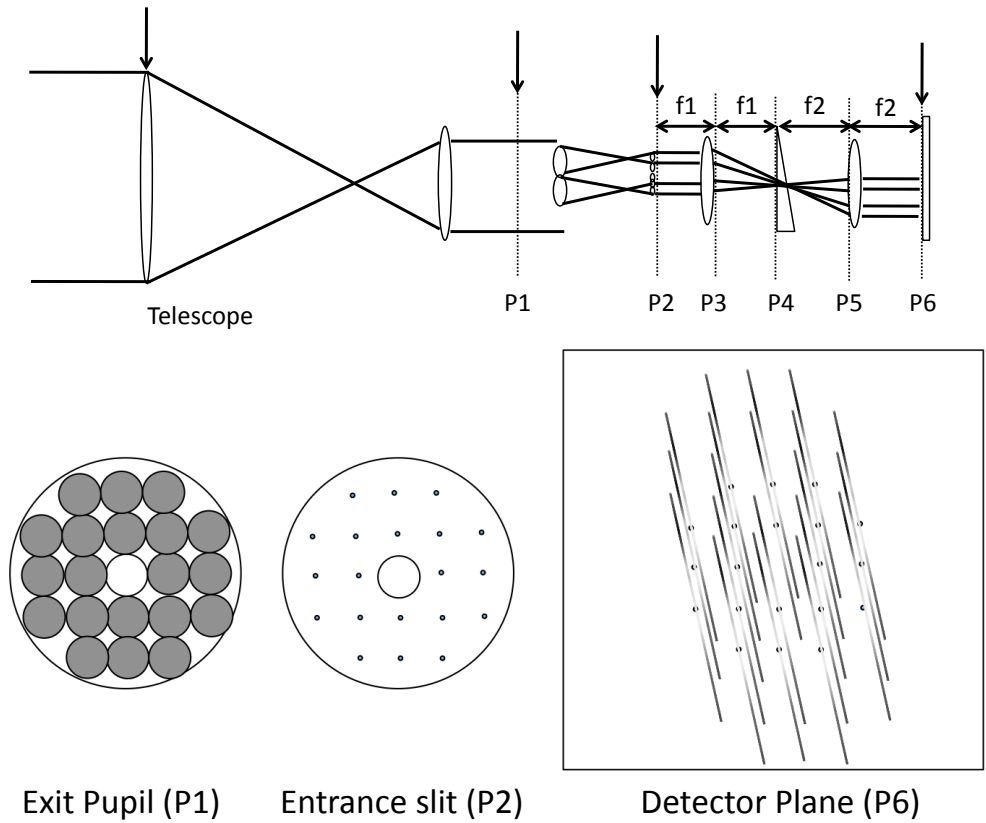}
\caption{Conceptual design of the densified pupil spectroscopy. The downward-pointing arrows represent the pupil planes that are an optical conjugate with the primary mirror. The cross-section view of each pupil plane is shown in the diagram beneath.  \label{fig1}}
\end{figure}

\begin{figure}
\epsscale{.80}
\plotone{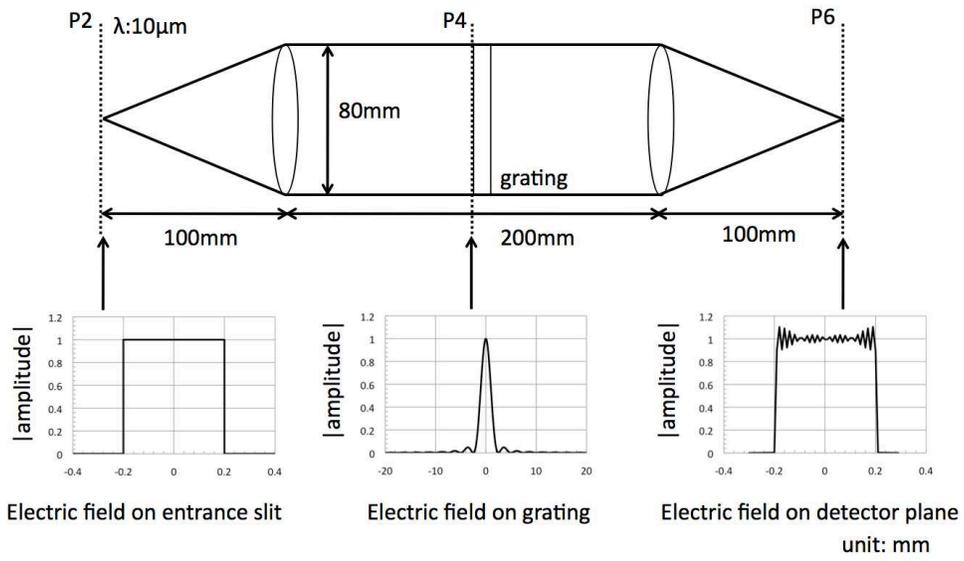}
\caption{Overall view of the spectrometer used for the calculation (upper panel) and the absolute values of the complex amplitudes on the entrance slit of the spectrometer (left-hand panel), on the dispersive element (center panel), and on the detector plane (right-hand panel). \label{fig2}}
\end{figure}

\begin{figure}
\epsscale{.80}
\plotone{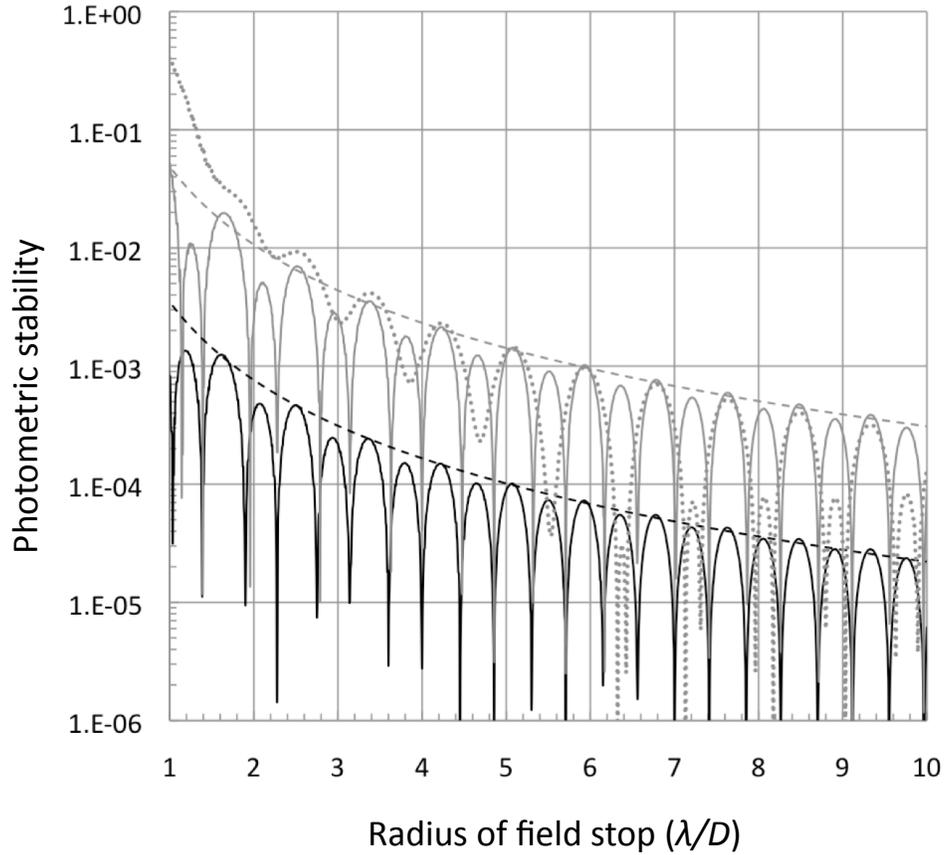}
\caption{Photometric stabilities as a function of the radius of the field stop put on the image plane for various pointing jitters. The black solid, gray solid, and gray dotted lines, respectively, show the photometric stabilities for the pointing jitters of $0.1\lambda/D$, $0.5\lambda/D$, and $1\lambda/D$. The black and gray dashed lines represent the approximated curves of the upper limit on the photometric stability for the pointing jitters of $0.1\lambda/D$ and more than $0.5\lambda/D$. \label{fig3}}
\end{figure}

\begin{figure}
\epsscale{.80}
\plotone{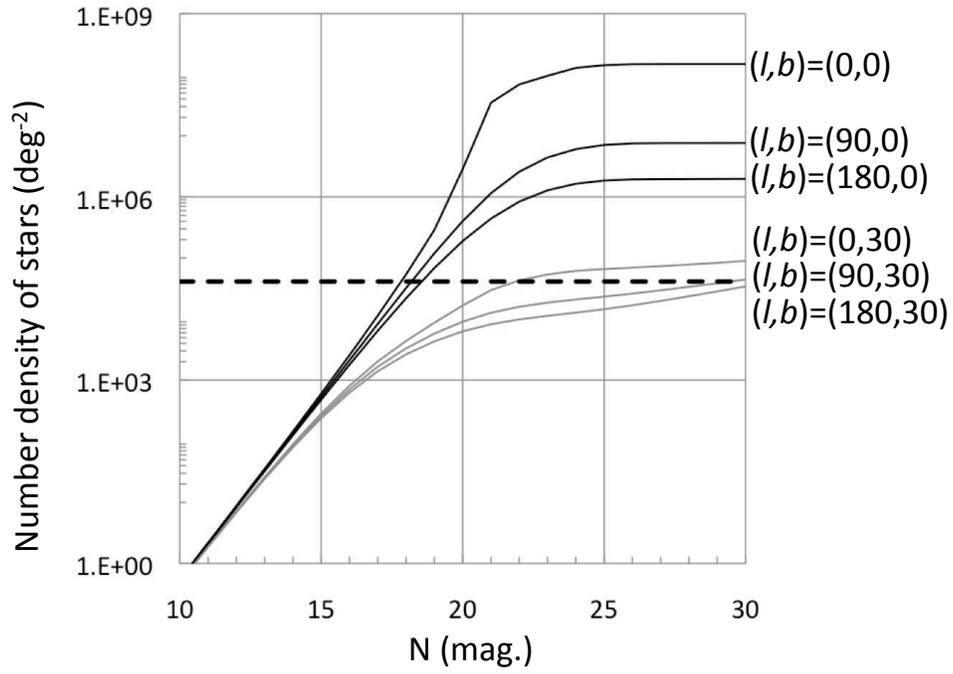}
\caption{Cumulative number densities of galactic stars in the N band for various galactic coordinates, $(l,b)=(0,0), (0,30), (90, 0), (90, 30), (180, 0)$, and $(180, 30)$. \label{fig3}}
\end{figure}

\begin{figure}
\epsscale{.80}
\plotone{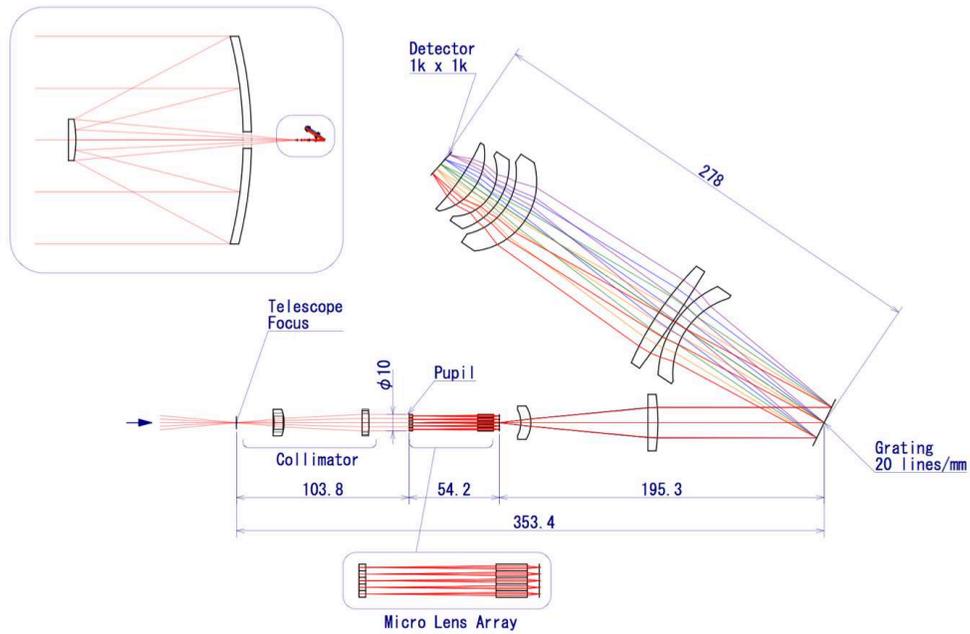}
\caption{Overall view of an optimized optical system for this concept. The upper left figure shows the optical system combined with a telescope with a 2.5$m$ diameter. The purple, blue, green, yellow, and red lines respectively represent the optical paths at 10$\micron$, 12$\micron$, 14$\micron$, 16$\micron$, 18$\micron$, and 20$\micron$. The material used in all the lenses is KRS5, through which mid-infrared light optically transmits. The scale unit is $mm$. \label{fig3}}
\end{figure}

\begin{figure}
\epsscale{.80}
\plotone{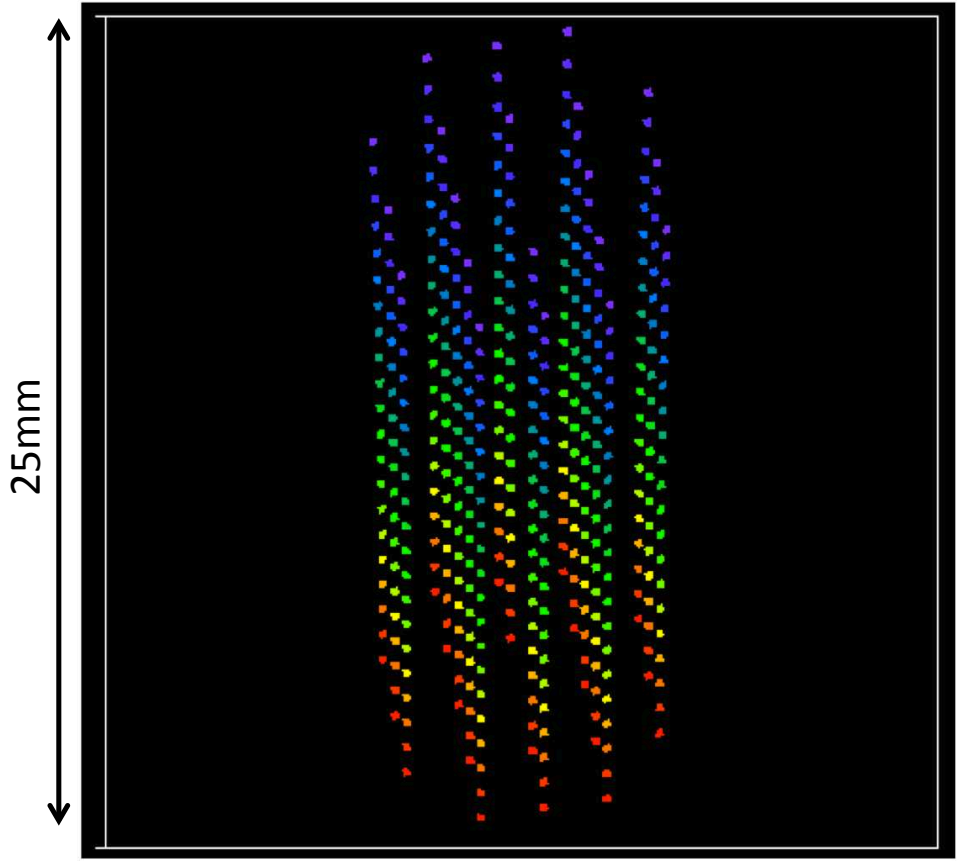}
\caption{Footprint of the detector. Twenty spectra are formed on the detector with a pixel format of 1000 x 1000 and a pitch of 25$\micron$. The color represents the wavelength, same as in Figure 3.  \label{fig4}}
\end{figure}

\begin{figure}
\epsscale{.80}
\plotone{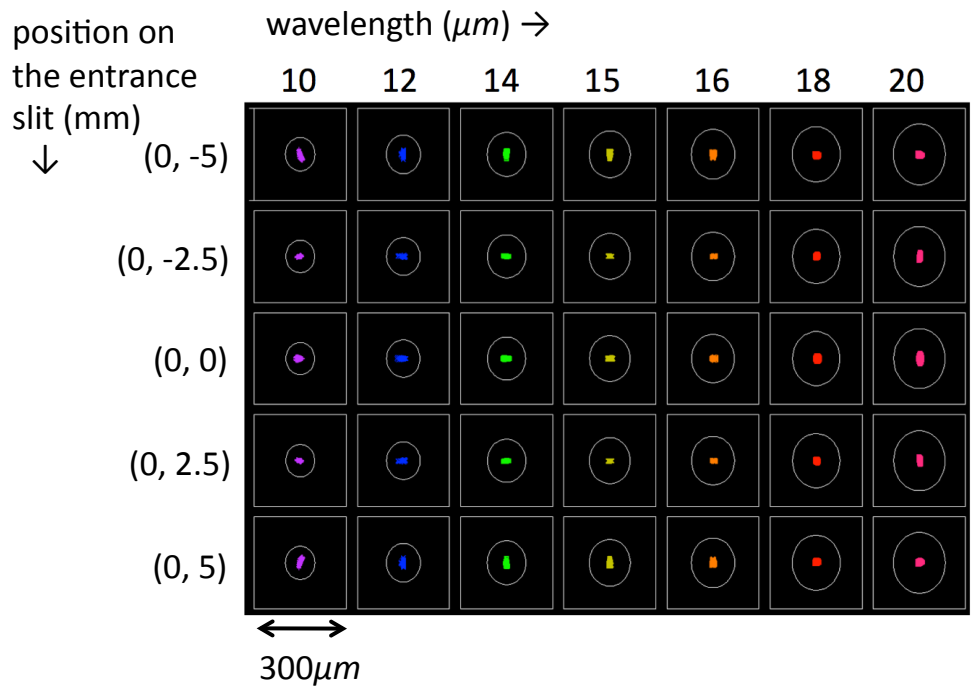}
\caption{Spot diagrams on the detector for various positions and wavelengths of the sub-pupils on entrance slit. \label{fig5}}
\end{figure}

\begin{figure}
\epsscale{.70}
\plotone{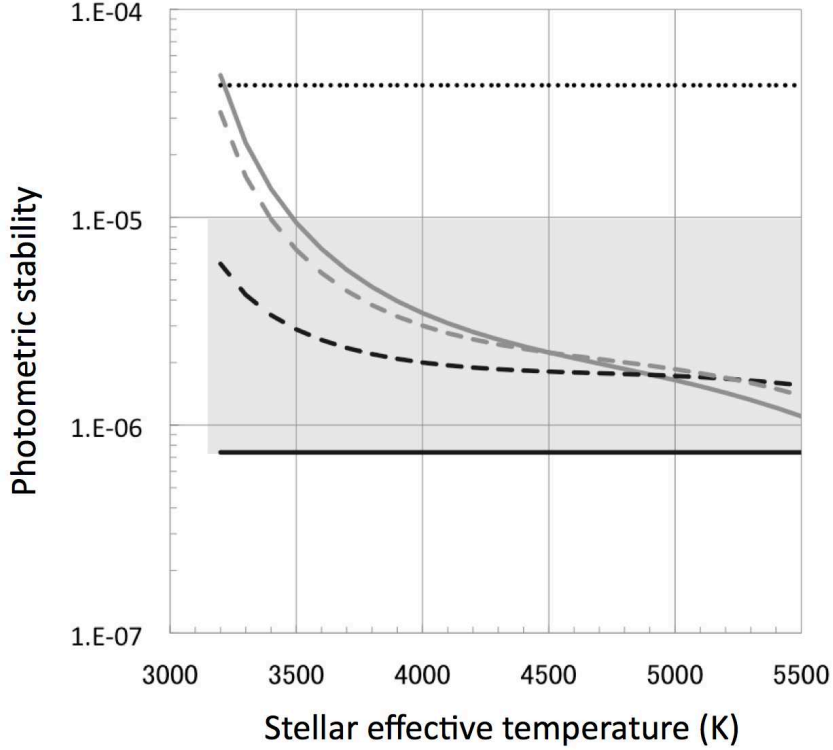}
\caption{Photometric stability of the system shown in Figure 5 and a requirement on 1-$\sigma$ detection of the secondary eclipse of a temperate super-Earth with an effective temperature 300$K$ and its atmosphere of 20$km$ scale height in the transmission spectroscopy at 10$\micron$. The black solid and dotted lines represent the 1-$\sigma$ instrumental systematic noises for the new concept and the conventional spectrometer caused by the telescope pointing jitter of 0.1 arc-second rms in total, respectively. The gray shaded area indicates the expected range of sum of the instrumental systematic noises including partial occultation of the target star with the field stop. The conventional spectrometer has 8 pixel sampling per one spectral channel in the same manner as the new system. The black dashed line is random noise including photon noise and detector noise. The gray solid line is the ratio of the thermal emission of a super-Earth with an effective temperature of 300$K$ to those of stars around effective temperatures of 3200$K$ to 5500$K$ at 10$\micron$. The gray dashed line shows requirement on 1-$\sigma$ detection of the atmospheric scale height of 20$km$, which corresponds to the thickness of the ozone layer and the height of the carbon dioxide layer in the Earth atmosphere, through transmission spectroscopy. \label{fig5}}
\end{figure}

\begin{table}
\begin{center}
\caption{Mean and worst values of image center movements on the detector plane in case of the telescope pointing jitter of $\pm$0.5 arc-second.\label{tbl-1}}
\begin{tabular}{crrrrrrrrrrr}
\tableline\tableline
Wavelength ($\micron$) & 10 & 12 & 14 & 16 & 18 & 20 \\
\tableline
Mean value (pixel)\tablenotemark{a} &0.0075 & 0.0070 & 0.0071  & 0.0074 & 0.0113 & 0.0162\\
Worst value (pixel)\tablenotemark{a} &0.013 & 0.016 & 0.016 & 0.016  &0.021  &0.036\\
\tableline
\end{tabular}
\tablenotetext{a}{Average and worst values of the image motions of the 20 sub-pupils on the detector plane}
\end{center}
\end{table}

\begin{table}
\begin{center}
\caption{Mean and worst values of the instrumental systematic errors caused by the telescope pointing jitter of $\pm$0.5 arc-second.\label{tbl-1}}
\begin{tabular}{crrrrrrrrrrr}
\tableline\tableline
Wavelength ($\micron$) & 10 & 12 & 14 & 16 & 18 & 20 \\
\tableline
Mean value (ppm)\tablenotemark{a} & 0.74 & 0.72 & 0.72  & 0.74 & 1.0 & 1.1\\
Worst value (ppm)\tablenotemark{a} &1.0 & 1.1 & 1.1 & 1.1  & 1.3  & 1.7\\
\tableline
\end{tabular}
\tablenotetext{a}{Average and worst values of the systematic errors for the 20 sub-pupils on the detector plane}
\end{center}
\end{table}

\begin{table}
\begin{center}
\caption{Simulation parameters.\label{tbl-3}}
\begin{tabular}{ll}
\tableline\tableline
Parameter & Value \\
\tableline
Telescope diameter & 2.5$m$\\
\tableline
Telescope pointing jitter & 0.5 arc-second\\
\tableline
Wavelength & 10$\micron$\\
\tableline
Spectral resolution at 10$\micron$ & 20\\
\tableline
Optical throughput & 0.2\\
\tableline
Total observing period & 3 years\\
\tableline
Field of view & 10 arc-second\\
\tableline
Detector dark current & 0.17$e^{-}/s/pixel$\\
\tableline
Detector readout noise & 14$e^{-}$\\
\tableline
Flat-fielding error & 0.07\%\\
\tableline
Distance of the system & 10$pc$\\
\tableline
Effective temperature of a super-Earth & 300$K$\\
\tableline
Radius of a super-Earth & 12600$km$\\
\tableline
Flux of exozodiacal light & 100$\mu Jy$\\
\tableline\tableline
\end{tabular}
\end{center}
\end{table}

\begin{table}
\begin{center}
\caption{Simulation parameters.\label{tbl-4}}
\begin{tabular}{llllll}
\tableline\tableline
Temperature & Radius & Mass & Semi-major axis & Transit duration & \# of eclipses per 1yr\\
\tableline
$K$ & $R_{\odot}$ & $M_{\odot}$ & $AU$ & Hours & \\
\tableline
3200 & 0.18 & 0.14 & 0.041 & 1.28 & 45.9 \\
\tableline
3300 & 0.26 & 0.25 & 0.062 & 1.68 & 32.8\\
\tableline
3400 & 0.33 & 0.34 & 0.083 & 2.11 & 24.6\\
\tableline
3500 & 0.39 & 0.41 & 0.10 & 2.55 & 19.2\\
\tableline
3600 & 0.45 & 0.47 & 0.13 & 2.97 & 15.6\\
\tableline
3700 & 0.49 & 0.52 & 0.15 & 3.37 & 13.0\\
\tableline
3800 & 0.53 & 0.56 & 0.17 & 3.75 & 11.1\\
\tableline
3900 & 0.57 & 0.60 & 0.19 & 4.10 & 9.6\\
\tableline
4000 & 0.60 & 0.63 & 0.21 & 4.42 & 8.5\\
\tableline
4100 & 0.62 & 0.65 & 0.23 & 4.75 & 7.5\\
\tableline
4200 & 0.64 & 0.67 & 0.25 & 5.05 & 6.7\\
\tableline
4300 & 0.66 & 0.69 & 0.26 & 5.31 & 6.1\\
\tableline
4400 & 0.68 & 0.70 & 0.28 & 5.59 & 5.5\\
\tableline
4500 & 0.69 & 0.72 & 0.30 & 5.82 & 5.1\\
\tableline
4600 & 0.71 & 0.73 & 0.32 & 6.08 & 4.7\\
\tableline
4700 & 0.72 & 0.74 & 0.34 & 6.33 & 4.3\\
\tableline
4800 & 0.73 & 0.76 & 0.36 & 6.55 & 4.0\\
\tableline
4900 & 0.75 & 0.77 & 0.39 & 6.83 & 3.7\\
\tableline
5000 & 0.76 & 0.78 & 0.41 & 7.14 & 3.4\\
\tableline
5100 & 0.78 & 0.80 & 0.44 & 7.43 & 3.1\\
\tableline
5200 & 0.80 & 0.82 & 0.47 & 7.77 & 2.9\\
\tableline
5300 & 0.82 & 0.84 & 0.50 & 8.17 & 2.6\\
\tableline
5400 & 0.85 & 0.86 & 0.53 & 8.64 & 2.4\\
\tableline
5500 & 0.88 & 0.88 & 0.57 & 9.19 & 2.2\\
\tableline\tableline
\end{tabular}
\end{center}
\end{table}

\clearpage




\end{document}